\documentclass{ws-procs975x65}

\begin{document}

\title{Phenomenology of Space Time Fluctuations}

\author{R. Aloisio\footnote{invited speaker}} 

\address{INFN - Laboratori Nazionali Gran Sasso \\
SS 17 bis, Assergi (AQ) Italy, \\ 
E-mail: roberto.aloisio@lngs.infn.it}

\author{P. Blasi}

\address{INAF - Osservatorio Astrofisico Arcetri \\
Largo E. Fermi 5, 50125 Firenze Italy\\ 
E-mail: blasi@arcetri.astro.it}

\author{A. Galante}

\address{Dipartimento di Fisica, Universit\`a di L'Aquila\\
Via Vetoio, 67100 Coppito (AQ) Italy\\ 
E-mail: angelo.galante@lngs.infn.it}

\author{A.F. Grillo}

\address{INFN - Laboratori Nazionali Gran Sasso \\
SS 17 bis, Assergi (AQ) Italy, \\ 
E-mail: aurelio.grillo@lngs.infn.it}


\maketitle

\abstracts{
Quantum gravitational effects may induce stochastic fluctuations in the
structure of space-time, to produce a characteristic foamy structure. 
It has been known for some time now that these fluctuations may have
observable consequences for the propagation of cosmic ray particles
over cosmological distances. While invoked as a possible explanation for 
the detection of the puzzling cosmic rays with energies in excess of the 
threshold for photopion production (the so-called super-GZK particles), we
demonstrate here that lower energy observations may provide 
strong constraints on the role of a fluctuating space-time structure.
We note also that the same fluctuations, if they exist, imply that 
some decay reactions normally forbidden by elementary conservation laws, 
become kinematically allowed, inducing the decay of particles that are seen 
to be stable in our universe. Due to the strength of the prediction, we are 
led to consider this finding as the most severe constraint on the classes of 
models that may describe the effects of gravity on the structure of space-time.
We also propose and discuss several potential loopholes of our approach, 
that may affect our conclusions. In particular, we try to identify the
situations in which despite a fluctuating energy-momentum of the
particles, the reactions mentioned above may not take place.
}

\section{Introduction}

In the last few years the hunt for possible minuscule violations of the
fundamental Lorentz invariance (LI) has been object of renewed interest,
in particular because it has been understood that cosmic ray physics has
an unprecedented potential for investigation in this field
\cite{kir,lgm,cam,colgla,noi,spain}. Some authors \cite{cam,colgla,berto}
have even invoked possible violations of LI as a plausible explanation to
some puzzling observations related to the detection of ultra high energy
cosmic rays (UHECRs) with energy above the so-called GZK feature \cite{gzk}, 
and to the unexpected shape of the spectrum of photons with super-TeV 
energy from sources at cosmological distances.
 
Both types of observations have in fact 
many uncertainties, either coming from limited statistics of very rare events,
or from accuracy issues in the energy determination of the detected 
particles, and most likely the solution to the alleged puzzles will come from 
more accurate observations rather than by a violation of fundamental 
symmetries.
 
For this reason, from the very beginning we proposed \cite{noi} that 
cosmic ray observations should be used as an ideal tool to constrain 
the minuscule violations of LI, rather than as evidence for the need 
to violate LI. The reason why the cases of UHECRs and TeV gamma rays 
represent such good test sites for LI is that both are related to physical
processes with a kinematical energy threshold, which is in turn very sensitive
to the smallest violations of LI. UHECRs are expected to suffer severe 
energy losses due to photopion production off the photons of the cosmic 
microwave background (CMB), and this should suppress the flux of particles 
at the Earth at energies above $\sim 10^{20}$ eV, the so called GZK feature. 

Present operating experiments are AGASA \cite{AGASA}
and HiRes \cite{Hires}, and they do not provide strong evidence either 
in favor or 
against the detection of the GZK feature \cite{demarco}. A substantial 
increase in the statistics of events, as expected with the Auger project
\cite{Auger}
and with EUSO \cite{EUSO}, 
should dramatically change the situation and allow to detect
the presence or lack of the GZK feature in the spectrum of UHECRs. These
are the observations that will provide the right ground for imposing a 
strong limit on violations of LI. For the case of TeV sources, the process
involved is pair production \cite{gamgam} of high energy gamma rays on
the photons of the infrared background. In both cases, a small violation
of LI can move the threshold to energies which are smaller than the 
classical ones, or move them to infinity, making the reactions impossible.
The detection of the GZK suppression or the cutoff in the gamma ray 
spectra of gamma ray sources at cosmological distances will prove that 
LI is preserved to correspondingly high accuracy \cite{noi}.

The recipes for the violations of LI generally consist of requiring an 
{\it explicit} modification of the dispersion relation of high energy 
particles, due to their propagation in the ``vacuum'', now affected by 
quantum gravity (QG). This effect is generally parametrized by introducing 
a typical mass, expected to be of the order of the Planck mass ($M_P$), 
that sets the scale for QG to become effective. 

However, explicit modifications of the dispersion relation are not really
necessary in order to produce detectable effects, as was recently
pointed out in Refs. \cite{ford,ng1,ng2,lieu} for the case of propagation 
of UHECRs. It is in fact generally believed that coordinate measurements 
cannot be performed with precision better than the Planck distance (time) 
$\delta x \geq l_P$, namely the distance where the metric of space-time 
must feature quantum fluctuations.
A similar line of thought implies that an uncertainty in the measurement
of energy and momentum of particles can be expected, according with the  
relation $\delta p \simeq \delta E \simeq p^2/M_P$. As discussed also in 
Refs. \cite{ng1,ng2} the apparent problem of super-GZK particles
may find a solution also in the context of this uncertainty approach.

We discuss here this appealing approach more in detail, by taking into 
account the effects of the propagation of CRs in the QG vacuum in the 
presence of the universal microwave background radiation. A fluctuating
metric implies that different measurements of the particle energy or 
momentum may result in different outcomes. Therefore it becomes important to
define the probability that the {\it measured} energy (momentum) of a particle 
is above some fixed value. Note that averaging
over a large number of measurements would yield the {\it classical} values
for the energy and momentum. The process of measurement mentioned above, 
during the propagation of particles over cosmological distances occurs at
each single interaction of the particle with the environment. At each
interaction vertex, the fluctuating energy/momentum of the particle is 
compared with the kinematic threshold for the occurrence of some physical 
process (in our case the photopion production).
A clear consequence of this approach is that particles with classical energy
below the standard Lorentz invariant threshold have a certain probability of 
interacting. In the same way, particles above the classical threshold have a 
finite probability of evading interaction. We show here that the most striking
consequences of the approach described above derive from low energy particles 
rather than from particles otherwise above the threshold for photopion 
production. 

However, the possibility of a fluctuating energy and momentum is mainly 
constrained by other processes that could arise. The fluctuations 
of energy and momentum are responsible, infact, for decaying processes 
otherwise impossible, typically prevented by energy and momentum conservation.
These decaying processes represent the most stringent test of the proposed 
model. In the present paper we will discuss these decaying processes, showing 
how they could arise. From a general point of view a particle 
propagating in a fluctuating vacuum acquires an energy dependent fluctuating 
effective mass (the fluctuating dispersion relations introduced in 
\cite{noi2}) which may be responsible for kinematically forbidden decay 
reactions to become kinematically allowed. 

If this happens, particles that are known to be stable would decay, provided 
no other fundamental conservation law is violated (e.g.: baryon number 
conservation, charge conservation). A representative example is that of the 
reaction $p\to p+\pi^0$, that is prevented from taking place only due to 
energy conservation. With a fluctuating metric, we find that if the initial 
proton has energy above a few $10^{15}$ eV, the reaction above can take place 
with a cross section typical of hadronic interactions, so that the proton 
would rapidly lose its energy. Similar conclusions hold for the 
electromagnetic process $p\to p+\gamma$.

The fact that particles that would be otherwise stable could
decay has been known for some time now \cite{gmestres,liberati} and
in fact it rules out a class of non-fluctuating modifications of the 
dispersion relations for some choices of the sign of the modification: 
the new point here is that it does not appear to be possible to fix the 
sign of the fluctuations, so that the conclusions illustrated above 
seem unavoidable. This result represents the most striking test of the 
fluctuating picture discussed in this paper and could in principle invalidate 
the basis of the proposed model itself.

The plan of the paper is the following: in \S 2 we discuss the effect of 
fluctuations on the propagation of high energy particles, setting also the
computational framework of the paper.
In \S 3, we discuss, mainly from the astrophysical point of view, the 
possibility of putting under experimental scrutiny some of the conclusions
reached in \S 2. In section \S 4 we will discuss the decays of stable 
particles induced by fluctuations. Finally in section \S 5 we argue that the 
comparison of our predictions with experimental 
data indicates a strong inconsistency, implying that the framework of 
quantum fluctuations currently discussed in most literature is in fact 
ruled out. The strength of this conclusion leads us to try to identify 
possible loopholes in our working assumptions. The ways to avoid the
dramatic effects of the fluctuating energy-momentum of a particle should
be mainly searched in the dynamics of Quantum Gravity.
These effects, in which our knowledge is poor to say the least, might 
forbid processes even when these processes are kinematically allowed
due to the fluctuations in the energy and momentum. 

\section{The effect of Space-Time fluctuations on the propagation
of high energy particles.}

While electroweak and strong interactions propagate through space-time,
gravity turns out to be a property of the space-time itself. This simple 
statement has profound implications in the quantization of gravity. Our 
belief that gravity can be turned into a quantum theory immediately implies 
that the structure of space-time has quantum fluctuations itself. 
Another way of rephrasing this concept is that space-time is expected to have 
a granular (or foamy) structure, where however the size of space-time cells 
fluctuates stochastically, thereby causing an intrinsic uncertainty in the 
measurements of space-time lengths, and indirectly of energy and momentum of 
a particle moving through space-time. The uncertainty appears on scales 
comparable with the Planck scale (the quantization scale of gravity).

It is generally argued that measurements of distances (times) smaller than 
the Planck length (time) are conceptually unfeasible, since the process of 
measurement collects in a Planck size cell an energy in excess of the Planck
mass, hence forming a black hole, in which information is lost.
This can be translated in different ways 
into an uncertainty on energy-momentum measurements \cite{ng1,ng2}. The
Planck length is a good estimate of the uncertainty in the De Broglie 
wave-length $\lambda$ of a particle with momentum $p$. Therefore 
$\delta \lambda \approx l_P$, and  $\delta p = \delta (1/\lambda) \approx 
(p^2 l_P)=(p^2/M_P)$.

Speculating on the exact characteristics of the fluctuations induced by
QG is beyond the scope of the present paper, and it would probably be
useless anyway, since the current status of QG approaches does not allow
such a kind of knowledge. We decided then to adopt a purely phenomenological
approach, in which some reasonable assumptions are made concerning the 
fluctuations in the fabric of space-time, and their consequences for the 
propagation of high energy particles are inferred. Comparison with 
experimental data then possibly constrains QG models.

Following \cite{ng1}, we assume that in each measurement:

\begin{itemize}

\item{the values of energy (momentum) fluctuate around their average values
(assumed to be the result theoretically recoverable for an infinite number
of measurements of the same observable): 
\begin{equation}
 E \approx  {\bar E} + \alpha \frac{\bar{E}^2}{M_P} 
\label{eq:Ebar}
\end{equation}
\begin{equation}
 p \approx  {\bar p} + \beta \frac{\bar{p}^2}{M_P} 
\label{eq:pbar}
\end{equation}
with $\alpha, \beta$ normally distributed variables and $p$ the modulus 
of the 3-momentum (for simplicity we assume rotationally invariant 
fluctuations);}

\item{the dispersion relation fluctuates as follows: 
\begin{equation}
P_\mu g^{\mu\nu} P_\nu = E^2-p^2 + \gamma \frac{p^3}{M_P}=m^2
\label{eq:PmuPmu}
\end{equation}
and $\gamma$ is again a normally distributed variable.}

\end{itemize}

Ideally, QG should predict the type of fluctuations introduced above, but,
as already stressed, this is currently out of reach, therefore we assume here 
that the fluctuations
are gaussian. Our conclusions are however not sensitive to this assumption:
essentially any symmetrical 
distribution with variance $\approx 1$, within a large factor, would
give essentially the same results.
Furthermore we {\it assume} that  $\alpha$, $\beta$ and $\gamma$ 
are uncorrelated random variables; again, this assumption reflects our 
ignorance in the dynamics of QG 

The fluctuations described above will in general derive from metric 
fluctuations of magnitude $\delta g^{\mu\nu} \sim h^{\mu\nu} \frac{l_P}{l}$ 
\cite{cam,ng2}. Our assumption reflects the fact that, while the magnitude of
the fluctuation can be guessed, we do not make any assumption on its 
tensorial structure $h^{\mu\nu}$.

Our interest will be now concentrated upon processes of the type

$$a+b\to c+d$$

where we assume that a kinematic threshold is present; in the realm of 
UHECR physics (a,b) is either ($\gamma, \gamma_{3K}$)
or ($p,\gamma_{3K}$) and (c,d) is ($e^+,e^-$) or ($N,\pi$).

To find the value of initial momenta for which the reaction occurs we write
down energy-momentum conservation equations and solve them with the
help of the dispersion relations, as discussed in detail in \cite{noi}.

The energy momentum conservation relations are (in the laboratory frame,
and specializing to the case in which the target (b) is a low energy
background photon for which fluctuations can be entirely neglected)
\begin{equation}
E_a+ \alpha_a \frac{E^2_a}{M_P} + \omega = 
E_c+ \alpha_c \frac{E^2_c}{M_P}+ E_d + \alpha_d \frac{E^2_d}{M_P}
\label{eq:disp1}
\end{equation}
\begin{equation}
p_a+ \beta_a \frac{p^2_a}{M_P} - \omega = 
p_c+ \beta_c \frac{p^2_c}{M_P}+ p_d + \beta_d \frac{p^2_d}{M_P}.
\label{eq:disp2}
\end{equation}
These equations refer to head-on collisions and collinear
reaction products, which is appropriate for threshold computations. Together
with the modified dispersion relations, these equations, after 
some manipulations, lead to a cubic equation for the initial momentum 
as a function of the momentum of one of products, and, after minimization, 
they define the threshold for the process considered. In figure 1 we 
report the distribution of thresholds in the $\approx 70 \%$ of cases
in which the solution is physical; in the other cases the kinematics 
does not allow the reaction.

This threshold distribution can be interpreted in the 
following way: a particle with energy above $\sim 10^{15}$ eV has essentially 
$70 \%$ probability of being above threshold, and therefore to be absorbed. 
In the other $30 \%$ of the cases the protons do not interact.

In (\ref{eq:disp1},\ref{eq:disp2}) the fluctuations are taken independently 
for each particle, which is 
justified as long as the energies are appreciably smaller than the Planck
energy. At that point it becomes plausible that different particles 
experience the same fluctuations, or more precisely fluctuations of the
same region of space-time. It is instructive to consider this case in some 
more detail: we introduce then the four-momenta (and dispersion 
relations) of {\it all} particles fluctuating in the same way. 
Specializing to proton interaction on CMBR, the equation which defines
the threshold $p_{th}$ is \cite{noi}:
\begin{equation}
\eta \frac{2 p_0^3}{(m_{\pi}^2+2m_{\pi}m_p)  M_P} 
\frac{m_{\pi}m_p}{(m_{\pi}+m_p)^2} 
\left ( \frac{p_{th}}{p_0} \right ) ^3   
+ \left ( \frac{p_{th}}{p_0} \right )   -1 = 0
\end{equation}
where $\eta$ is a gaussian variable with zero average and variance
of the order of (but not exactly equal to) one, and $p_0$ is
the L.I. threshold (GZK). The threshold is the positive solution of
this equation.

\begin{figure}[ht]
\centerline{\epsfxsize=3.8in\epsfbox{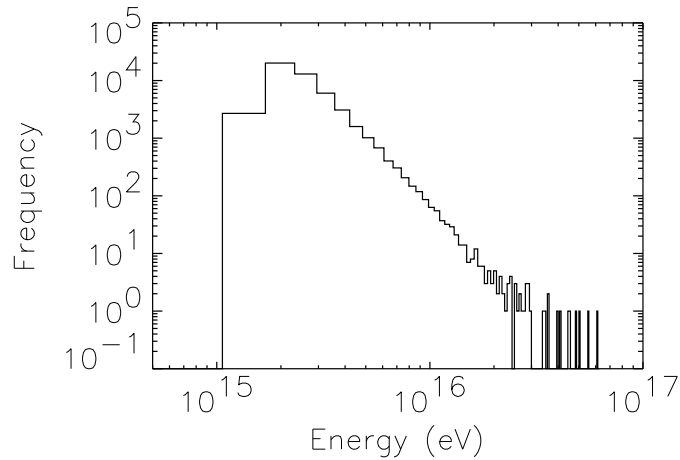}}   
\caption{Threshold distribution for $p \gamma_{3^oK} \to 
N \pi$. In the $30\%$ of cases the reaction is not allowed.}
\end{figure}
\eject

The coefficient of the cubic term is very large, of the 
order of $10^{13}$ in this case, so that unless $\eta$ 
is $O(10^{-13})$,
we can write, neglecting pion mass
\begin{equation}
p_{th}\approx p_0 \left ( \frac{m_p^2 M_P}{\eta p_0^3} 
\right )^{\frac{1}{3}}.
\label{eq:thre}
\end{equation}
When $\eta$ becomes negative, the above equation has no 
positive root; this happens essentially in $50 \%$ of the cases.
Since the gaussian distribution is  flat in a small interval 
around zero, the distribution of thresholds for positive $\eta$ 
peaks around the 
value for $\eta \approx 1$, meaning that the threshold moves almost 
always down to a value of $\approx 10^{15}$ eV \cite{noi}; 
essentially the same result holds for fluctuations affecting only  the 
incident (highest energy) particle.
For {\it independent} fluctuations of final momenta,
the asymmetry in the probability 
distribution of allowed thresholds arises from the fact that 
even
exceedingly small negative values of the fluctuations lead 
to unphysical solutions. 

Building upon our findings, we now apply the same calculations to 
the case of UHECR protons propagating on cosmological distances.
An additional ingredient is needed to complete the dynamics of the
process of photopion production, namely the cross section. The rather
strong assumption adopted here is that the cross section remains the
same as the Lorentz invariant one, provided the reaction is 
{\it kinematically} allowed. This implies that the interaction lengths
remain unchanged.

In order to assess the situation of UHECRs, we first consider the 
case of particles above the threshold for photopion production in 
a Lorentz invariant world. According with eqs. (\ref{eq:disp1},
\ref{eq:disp2}),
in this case particles have a probability of $\approx 30 \%$ of being
not kinematically allowed to interact inelastically with a photon in the 
CMBR. Therefore, if our assumption on the invariance of the interaction 
length is correct, then each proton is still expected to make photopion
production, although with a slightly larger pathlength. 

The situation is however even more interesting for particles that are
below the Lorentz invariant threshold for the process of photopion 
production. If the energy
is below a few $10^{18}$ eV, a galactic origin seems to be in good 
agreement with measurements of the anisotropy of cosmic ray arrival 
directions \cite{agasa_a,fly_a}. We will not consider these energies 
any longer. On the other hand, at energies in excess of $10^{19}$ eV,
cosmic rays are believed to be extragalactic protons, mainly on the ground of
the comparison of the size of the magnetized region of our Galaxy and
the Larmor radius of these particles. 
We take these pieces of information as the basis 
for our line of thought. If the cosmic rays observed in the energy range 
$E> 10^{19} \rm{eV}$ are extragalactic protons,
then our previous calculations apply and we may expect that these particles 
have a $\sim 70\%$ probability of suffering photopion 
production at each interaction with the CMB photons, even if their energy 
is below the classical threshold for this process. Note that the pathlength 
associated with the process is of the order of the typical pathlength for 
photopion production (a few  tens of Mpc), therefore we are here discussing 
a dramatic process in which the absorption length of particles drops from 
Gpc, which would be pertinent to particles with energy below 
$\sim 10^{20}$ eV in a Lorentz invariant world, to several Mpc, with a 
corresponding suppression of the flux. 
What are the consequences for the observed fluxes of cosmic rays? 
The above result implies that {\it all} protons with $E>10^{15}$ eV
are produced within a radius of 
several tens of Mpc, and above this energy there is no dramatic change 
of pathlength with energy. 
There is no longer anything
special about  $E \sim 10^{20}$ eV, and 
any mechanism invoked to explain the flux of super-GZK particles must
be at work also at lower energies. 

The basic situation remains the same in the case of pair production as the
physical process under consideration. For a source at cosmological distance,
a cutoff is expected due to pair production off the far infrared background
(FIR) or the microwave background. Using the results in \cite{noi} we expect
that the modified thresholds are a factor $0.06$ ($0.73$) lower than the 
Lorentz invariant ones for the case of interaction on the CMBR (FIR).
There is also a small increase in the pathlengths above the threshold,
which would appear exponentially in the expression for the flux. Therefore
there are two effects that go in opposite directions: the first moves the
threshold to even lower energies, and the second increases the flux of
radiation at Earth because of the increase of the pathlength. 
It seems that geometry fluctuations do not provide an immediate explanation
of the possible detection of particles in excess of the expected ones from
distance sources in the TeV region. In any case the experimental evidence 
for such an excess seems at present all but established.

\section{Astrophysical observations}

As discussed in the previous section fluctuations in the space-time metric 
may induce a violation of Lorentz invariance that changes the thresholds for 
the photopion production of a very high energy proton 
off the photons of the CMBR, or for the pair production of a high energy
gamma ray in the bath of the FIR or CMBR photons.

For the case of UHECRs interacting with the CMBR, we obtained a picture that 
changes radically our view of the effect of QG on this phenomenon, as 
introduced in previous papers: not only particles with energy above 
$\sim 10^{20}$ eV are affected by the fluctuations in space-time, but also 
particles with lower energy, down to $\sim 10^{15}$ eV seem to be affected
by such fluctuations. In fact the latter, as a result of a fluctuating 
space-time, may end up being above the threshold for photopion production, 
so that particles may suffer significant absorption. Our conclusion is that 
all particles with energy in excess of $\sim 10^{15}$ eV eventually detected 
at Earth would be generated at distances comparable with the pathlength 
for photopion production ($\sim 100$ Mpc). 
A consequence of this is that there is no longer anything special 
characterizing the energy $\sim 10^{20}$ eV. 

Since the conclusion reached in the previous section is quite strong, it is 
important to summarize in detail some tests that may allow to understand 
whether the current or future astrophysical observations are compatible with 
the scenario discussed in this paper.

a)  Future experiments \cite{Auger,EUSO} 
dedicated to the detection of UHECRs will provide a
substantial increase in the statistics, so that the spectral features of 
the UHECRs in the energy region $E>10^{19}$ eV can be resolved, and
 further indications on the nature of primaries and their 
possible extragalactic origin will be obtained. In particular
the present possible disagreement between  
AGASA \cite{agasa_f} and HiRes \cite{Hires_a} will be clarified.

One should also keep in mind that an evaluation of the expected flux in 
terms of sources distributed as normal galaxies is
in contradiction with AGASA data by an amount ranging from $2$ to $6 \sigma$
depending on the assumed source spectrum \cite{blanton}. 
Since the nature of the sources is not known, it is not
clear if their  abundance within the absorption pathlength
is sufficient to explain the observed flux in presence of space-time 
fluctuations, nor if they can induce observable anisotropies.

In any case, in a Lorentz invariant framework a suppression in the flux 
at $\sim 10^{20}$ eV is expected. If such a feature is 
unambiguously detected in the UHECR spectrum, no much room would be left for
the fluctuations of space-time discussed in this paper, since in this scenario
nothing special happens around $10^{20}$ eV.
In quantitative terms \cite{noi} 
this would imply  a phenomenological bound on $l_P$ 
now interpreted as a parameter: 
 $l_P < 10^{-46}$ cm instead of 
$l_P \approx 10^{-33}$ cm; in other words, only fluctuations with 
variance $\approx 10^{-13}$, instead of $1$, would be allowed
\footnote{ Alternatively, 
one can assume a more general form of fluctuations, i.e. $\delta E \approx 
E (E/M_P)^{\alpha}$ and similar for momentum and dispersion relations 
\cite{tom}. In this case the basic conclusions reached here remain unchanged.}

b) According with our findings, all particles with energy in excess of 
$\sim 10^{15}$ eV lose their energy by photopion production on cosmological
spatial scales, as a result of the metric fluctuations. This energy ends up 
mainly in gamma rays, neutrinos and protons. The protons pile up in the
energy region right below $\sim 10^{15}$ eV. The gamma ray component
actually generates an electromagnetic cascade that ends up contributing 
low energy gamma rays, in the energy band accessible to instruments like
EGRET \cite{EGRET} and GLAST \cite{GLAST}. 
This cascade flux cannot be larger than the measured 
electromagnetic energy density in the same band
$\omega_{cas}^{exp}=10^{-6}~eV/cm^{3}$ \cite{EGRET}. The cascade flux in our 
scenario can be estimated as follows. Let $\Phi(E)=\Phi_0 (E/E_0)^{-\gamma}$
be the emissivity in UHECRs ($\rm{particles}/cm^3/s/GeV$). Let us choose the
energy $E_0=10^{10}$ GeV and let us normalize the flux to the observations
at the energy $E_0$. The total energy going into the cascade can be shown 
to be $$\omega_{cas} \approx \frac{5\times 10^{-4}}{\gamma-2}~~ 
x_{min}^{2-\gamma} ~~\xi ~~eV ~cm^{-3},$$ 
where $\xi$ is the fraction of energy going into gamma rays in each 
photopion production, and $x_{min}=(E_{th}/E_0)=10^{-4}$ for $E_{th}=
10^{15}$ eV. It is easy to see that, for $\gamma=2.7$, the cascade bound 
is violated unless $\xi \ll 10^{-3}$.

One note of warning has to be sent concerning the development of the 
electromagnetic cascade: the same violations of LI discussed here affect 
other processes, as stressed in the paper. For instance pair production 
and pion decay are also affected by violations of LI \cite{amepai}. 
Therefore the possibility that the cascade limit is exceeded concerns 
only those scenarios of violations of LI that do not inhibit appreciably 
pair production and the decay of neutral pions.

The protons piled up at energies right below $10^{15}$ eV, would be a 
nice signature of this scenario, but it seems difficult to envision a 
way of detecting these remnants. In fact, even a tiny magnetic field
on cosmological scales would make the arrival time of these particles to
Earth larger than the age of the universe. Moreover, even assuming an exactly
zero extragalactic magnetic field, these particles need to penetrate the 
magnetic field of our own Galaxy and mix with the galactic cosmic rays, 
making their detection extremely problematic if not impossible.

Clearly a more detailed flux computation, taking into account propagation of 
primaries as well as generation and propagation of the secondaries is needed 
in order to assess in a more quantitative way observable effects of possible 
metric fluctuations on UHECRs.

Let us conclude this section sending a note of warning concerning 
Eq. (\ref{eq:thre}), in this expression the dependence on the CMB photon
energy is washed out by the approximation done (we have neglected the pion 
mass). From the physical point of view this corresponds to the appearence 
of an effective mass (momentum dependent) of the proton due to the effect 
of fluctuations. The effective mass of the proton may be responsible 
for the decay of this particle. As we will discuss in the next section 
the possibility of a decaying proton is a very stringent test for the 
fluctuations picture much powerful than the astrophysical observations 
discussed in the present section. 

\section{Decay of stable particles}

Let us discuss in this section the most striking test of the models 
that predict energy and momentum fluctuations. We will discuss here 
the possibility that these fluctuations may induce particles decays otherwise 
impossible. This possibility, already discussed in the framework of 
non-fluctuating modifications of the dispersion relation 
\cite{gmestres,liberati}, could in principle rule out the models with 
fluctuations. In this section we will discuss the basic features of the
decays, leaving a detailed discussion of the implications and possible way
out to the next section.

We will consider three specific decay channels, that illustrate
well, in our opinion, the consequences of the quantum fluctuations introduced
above. We start with the reaction 
$$p\to p + \pi^0$$
and we denote with $p$ ($p'$) the momentum of the initial (final) proton, and
with $k$ the momentum of the pion. Clearly this reaction cannot take place 
in the reality as we know it, due to energy conservation. However, since 
fluctuations have the effect of emulating an effective mass of the particles, 
it may happen that for some realizations, the effective mass induced to the 
final proton is smaller than the mass of the proton in the initial 
state, therefore allowing the decay from the kinematical point of view. Since
no conservation law or discrete symmetry is violated in this reaction, it 
may potentially take place. For the sake of clarity, it may be useful to 
invoke as an example the decay of the $\Delta^+$ resonance, which is 
structurally identical to a proton, but may decay to a proton and a pion
according to the reaction $\Delta^+\to p+\pi^0$, since its mass is larger
than that of a proton. From the physical point of view, the effect of the
quantum fluctuations may be imagined as that of {\it exciting} the proton,
inducing a mass slightly larger than its own (average) physical mass.

Following the discussion of the previous sections we expect to find that for 
momenta above a given threshold, depending on the value of the random 
variables, the decay may become kinematically allowed. In general, the 
probability for this to happen has to be calculated  numerically from the 
conservation equations supplemented by the dispersion relations \cite{noi3}. 

Although a full calculation is possible, it is probably more instructive
to proceed in a simplified way, in which only the fluctuations in
the dispersion relation of the particle in the initial state are taken
into account. Neglecting the corresponding fluctuations in the final state
should not affect the conclusions in any appreciable way, unless the
fluctuations in the initial and final states are correlated (we will return 
to this possibility at the end of section \S 5).

In this approximation, the threshold for the process of proton decay to a 
proton and a neutral pion can be written as follows (neglecting corrections 
to order higher than $p/M_P$):

\begin{equation}
\gamma \frac{2 p_{th}^3}{M_P} 
-2 m_{\pi} m_p - m_{\pi}^2 =0,
\end{equation}
with solution
\begin{equation}
p_{th}= \left (\frac{(2m_p m_{\pi} +  m_{\pi}^2 )M_P}{2 \gamma } 
\right )^{\frac{1}{3}}.
\end{equation}
For negative values of $\gamma$, the above equation has no positive
root; this happens in $50\%$ of the cases. Since the gaussian distribution 
is essentially flat in a small interval around zero, the distribution of 
thresholds for positive $\gamma$ ({\it i.e. } in the remaining 50 $\%$ of the
cases) peaks around the value for $\gamma \approx 1$, meaning that the 
threshold moves almost always down to a value of $\approx 10^{15}$ eV 
\cite{noi,noi2}; essentially the same result holds for generic fluctuations 
({\it i.e.} not confined to the dispersion relations) affecting only the 
incident particle, namely the one with the highest energy \cite{noi3}.

The reason why the effects of fluctuations are expected to occur at such 
low energies is that starting from that energy region the fluctuation term
becomes comparable with the rest mass of the particle. In fact the same 
concept of rest mass of a particle may lose its traditional meaning at 
sufficiently high energies \cite{lieu}.

It can be numerically confirmed that {\it independent} fluctuations of  
momenta (and/or of the dispersion relations) of the decay products are more 
likely to make the decay easier rather than more difficult, due to the non 
linear dependence of the threshold on the strength of fluctuations:
the probability that the decay does not take place is in fact $\approx 30 \%$. 
In the remaining cases, the decay will occur if the momentum of the initial 
proton is larger than $p_{th}$\cite{noi3}. The distribution of $p_{th}$ 
is essentially identical to the one reported in \S 2 for the photopion 
production.

All the discussion reported so far remains basically unchanged if similar
reactions are considered. For instance the reaction $p \to \pi^+ n$ is
kinematically identical to the one discussed above. For all these reactions,
we expect that once they become kinematically allowed, the energy loss
of the parent baryon is fast. For the case of nuclei, all the decays that do
not change the nature of the nucleon leave (A,Z) unchanged, so we do
not expect any substantial blocking effect in nuclei.

Another reaction that may be instructive to investigate is the spontaneous 
pair production from a single photon, namely \cite{noi3}
$$\gamma \to e^+ e^-.$$
In this case, following the calculations described above, we obtain the 
following expression for the threshold:
\begin{equation}
p'_{th}= \left (\frac{4 m_e^2 M_P}{2 \gamma' }, 
\right )^{\frac{1}{3}}
\end{equation}
and $p'_{th}$ is of the order of $10^{13}$ eV.
Again, if the reaction becomes kinematically allowed, there does not seem 
to be any reason why the reaction should not take place with a rate 
dictated by the typical cross section of electromagnetic interactions. 

Finally, we propose a third reaction that in its simplicity may represent
the clearest example of reactions that should occur in a world in which 
quantum fluctuations behave in the way described above. Let us consider
a proton that moves in the vacuum with constant velocity, and let us consider
the elementary reaction of spontaneous photon emission. In the Lorentz 
invariant world the process of photon emission
is known to happen only in the presence of an external field that 
may provide the conditions for energy and momentum conservation. However, 
in the presence of quantum fluctuations, one can think of the gravitational
fluctuating field as such an external field, so that the particle can in fact 
radiate a photon without being in the presence of a nucleus or some other 
external recognizable field. The threshold for this process, calculated
following the above procedure, is 
\begin{equation}
p_{th}'' \approx \left (\frac{m^2 M_P \omega}{\gamma''}
\right )^{\frac{1}{4}},
\end{equation} 
where $\omega$ is the energy of the photon. This threshold approaches zero 
when $\omega \to 0$: for instance, if $\omega=1$ eV, then $p_{th} \approx 300$ 
GeV for protons and $p_{th} \approx 45$ GeV for electrons. In other words there
should be a sizable energy loss of a particle in terms of soft photons. 
This process can be viewed as a sort of bremsstrahlung emission of a charged
particle in the presence of the (fluctuating) vacuum gravitational potential.

Based on the arguments provided in this section, it appears that all particles
that we do know are stable in our world, should instead be unstable at 
sufficiently high energy, due to the quantum fluctuations described above. 
In the next section we will take a closer look at the implications of the 
existence of these quantum fluctuations, and possibly propose some plausible 
avenues to avoid these dramatic conclusions.

\section{Discussion and Outlook}

If the decays discussed in the previous section could take place, our
universe, at energies above a few PeV or even at much lower energies 
might be unstable, nothing like what we actually see.
The decays $$nucleon\to nucleon+\pi$$ would start to be kinematically allowed 
at energies that are of typical concern for cosmic ray physics, while 
the spontaneous emission of photons in vacuum might even start playing
a role at much lower energies, testable in laboratory experiments. 
Without detailed calculations of energy loss rates it is difficult to 
assess the experimental consequences of this process. 

For the nucleon decay, the situation is slightly simpler if we assume
that the quantum fluctuations affect only the kinematics but not the
dynamics, an assumption also used in in the photopion production study 
\cite{noi2}. In this case one would expect 
the proton to suffer the decay to a proton and a pion on a time scale of the
same order of magnitude of typical decays mediated by strong interactions.
This would basically cause no cosmic ray with energy above $\sim 10^{15}$
eV to be around, something that appears to be in evident contradiction 
with observations 
\footnote{From a phenomenological point of view, consistency 
with experiments would require either that the variance of the fluctuations 
considered above is ridiculously small ($<10^{-24}$) or, allowing more generic
fluctuations $\Delta l \propto l_P (l_P/l)^{\alpha}$, that a fairly large value
for $\alpha$ should be adopted \cite{noi2}.}. 

In the following we will try to provide a plausible answer to these three very 
delicate questions:
\begin{enumerate}
\item If the particles were kinematically allowed to decay, and there
were no fundamental symmetries able to prevent the decay, would it 
take place? 

\item Is the form adopted for the quantum fluctuations correct and if so,
how general is it?

\item If in fact the form adopted for the fluctuations is correct, how
general and unavoidable is the consequence that (experimentally)
unobserved decays should take place?
\end{enumerate}

Although the result that particles are kinematically allowed to
decay is fairly general, the (approximate) lack of relativistic
invariance forbids the computation of life-times \footnote{In fact
life-times can be in principle estimated in approaches in which it is
possible to make transformations between frames \cite{lieu,dsr,jap},
despite the lack of LI.}.
Two comments are in order: first, the phase space for the decays described
above, as calculated in the laboratory frame, is non zero and in fact 
it increases with the momentum of the parent particle. The effect of 
fluctuations can be seen as the generation of an effective (mass)$^2 
\propto p^3/M_P$. A similar effect, although in a slightly different 
context, was noted in \cite{colgla}.

Second, we do not expect dynamics to forbid the reactions:
one must keep in mind that we are considering very small effects, at
momenta much smaller than the Planck scale. For instance the gravitational
potential of the vacuum fluctuations is expected to move quarks in a
proton to excited levels, not to change its content, nor the properties 
of strong interactions. 

There is a subtler possibility, which must be taken very seriously in our
opinion, since it might invalidate completely the line of thought illustrated
above, namely that the quantum fluctuations of the momenta of the particles
involved in a reaction occur on time scales that are enormously smaller than 
the typical interaction/decay times. This situation might resemble the
so called Quantum Zeno paradox, where continuously checking for the
decay of an unstable particle effectively impedes its decay. 
This possibility is certainly worth a detailed study, that would however
force one to handle the intricacies of matter in a Quantum Gravity regime. 
We regard this possibility as the most serious threat to the validity of 
the arguments in favor of quantum fluctuations discussed in this paper 
and in many others before it. 

Let us turn out attention toward the question about the correctness and
generality of the form adopted for the momentum fluctuations. It is 
generally accepted that the geometry of space-time suffers profound 
modifications at length (time) scales of the order of the Planck
length (time), and that this leads to the emergence of a minimum
measurable length. This may be reflected in a non commutativity of 
space-time and in a generalized form of the uncertainty principle.

The transition from uncertainty in the length or time scales to uncertainty
in momenta of particles is undoubtedly more contrived and deserves some
attention. The expressions in Eqs. (\ref{eq:Ebar},\ref{eq:pbar}) and 
(\ref{eq:PmuPmu}) have been 
motivated in various ways \cite{ford,ng1,noi2,lieu,camacho} in previous
papers. For instance, the condition $\Delta l \ge l_P$ seems to imply the 
following
constraint on wavelengths $\Delta \lambda \ge l_P$, otherwise it would be 
possible to design an experimental set-up capable of measuring distances with 
precision higher than $l_P$. Therefore $\Delta p \propto \Delta (\lambda^{-1})
\propto  l_P p^2$. Similar arguments have been proposed, all based to some
extent on the de Broglie relation $p \propto \lambda^{-1}$. 

There is certainly no guarantee that the de Broglie relation continues
to keep its meaning in the extreme conditions we are discussing, in 
particular in models in which the coordinates and coordinate-momentum 
commutators are modified with respect to standard quantum mechanics and 
the representation of momentum in terms of coordinate derivatives generally 
fails. For instance in a specific (although non-relativistic) example 
\cite{kempf} the existence of a minimum length is shown to imply that
\begin{equation}
p = \frac{2}{\pi l_P} \tan \left ( \frac{\pi l_P}{2 \lambda} 
\right ).
\end{equation}
In other words, the de Broglie relation may be modified in such a way that
a minimum wavelength corresponds to an unbound momentum.
Notice, however, that we are considering here the effects of these
modifications at length scales much larger than the Planck scale, where the
correction is likely to be negligible. In general, if $p \propto
\lambda^{-1}g(l_P/ \lambda)$ then $\Delta p \propto l_P  p^2 +
p ~O(l_P^2 p^2)$. Hence, we do not expect that the result shown in the
previous Section is appreciably modified.

Last but not least we notice that the fluctuations in the dispersion 
relations can be easily derived from fluctuations of the (vacuum) metric in 
the form given in \cite{camacho}:
\begin{equation}
ds^2 = (1+\phi)dt^2-(1+\psi)d{\bf r}^2
\end{equation}
where $ \phi, ~\psi $ are functions of the position in space-time. 

The fluctuations of the dispersion relation, Eq. (\ref{eq:PmuPmu}), follow 
if $\phi \ne \psi$ ({\it i.e.} non conformal fluctuations), assuming at 
least approximate validity of the de Broglie relation; if $\phi=\psi$ a 
much milder modification (O($p m^2/M_P$)) follows.

Having given plausibility arguments in favor of the form adopted for the
fluctuations, at least for the case of non conformal fluctuations, we are 
left with the goal of proving an answer to the last question listed above, 
namely does a decay actually occur once it is kinematically allowed?
Certainly the answer is positive if one continues to assume momentum and
energy conservation, and modifications of these conservation laws with 
random terms of order $O(p^2/M_P)$ do not change this conclusion.
The question then is whether we are justified in assuming energy and momentum
conservation in the form used above. For instance, in the so-called Doubly
Special Relativity (DSR) \cite{dsr}, theories and in general in models with
deformed Poincare' invariance, the conservation relations may be
modified in a non trivial, non additive and non abelian way. For instance, 
in the case of proton decay considered above, momentum conservation
may read as \cite{dsr,jap}
\begin{equation}
{\bf p}_p \approx {\bf p}'_{p}+(1+l_P E'_p) {\bf p}_{\pi}\quad\quad
{\rm or}
\quad\quad 
{\bf p}_p \approx {\bf p}'_{\pi}+(1+l_P E'_{\pi}) {\bf p}_p.
\end{equation}
This certainly makes the probability of being above threshold smaller.
However in order to qualitatively modify our results this probability
should be in fact vanishingly small. For the case of {\it low} energy 
cosmic rays, this probability should be of the order of a typical decay
time divided by the residence time of cosmic rays (mostly galactic at
these energies) in our Galaxy. 

We are led to conclude that allowing for modifications of the conservation 
relations does not appear to improve the situation to the point that  
the strong conclusions derived in the previous section can be avoided.
In the same perspective, cancellation between fixed modifications 
of the dispersion relation and fluctuations (of the same order of 
magnitude) does not seem a viable way to proceed.

It is important however to notice that we have considered the above 
fluctuations as independent. In a full theory one should take into account
possible correlations between fluctuations. The effect of correlations is
very important because it pushes to higher energies the fluctuation scale
of the particle momentum (energy). Let us discuss in more detail this point.
Quantum fluctuations of the momenta of the particles involved in a reaction
occour on time scales that are much smaller than the typical interaction 
time. Particles during the interaction time experience a large number of 
fluctuations, typically

$$ N=\frac{\tau}{\tau_P}=\frac{1}{p\tau_P}=\frac{M_P}{p}~, $$
where we have used $\tau\sim 1/p$ for the interaction time scale and 
$\tau_P\sim 1/M_P$ for the fluctuation time scale. Assuming independent
fluctuations of energy and momentum the fluctuation variance $\sigma$ will be

$$ \sigma^2 = \frac{p^3}{M_P\sqrt{N}}=\frac{p^3}{M_P} 
\left (\frac{p}{M_P} \right )^{1/2}~, $$
and the fluctuation variance becomes of the order of the proton mass 
$\sigma\simeq m_p$ already at momentum $p\simeq 10^{17}$ eV. 
In this case the situation resembles as discussed above and, for instance, 
the decaying of the proton arises already at lower energies. Let us consider 
now the case in which there is some degree of correlation in the momentum 
(energy) fluctuations. In this case the fluctuation variance $\sigma$ will be

$$ \sigma^2 = \frac{p^3}{M_P N^{\alpha}}=\frac{p^3}{M_P} 
\left (\frac{p}{M_P} \right )^{\alpha}~, $$
where we have introduced the exponent $\alpha>1/2$ that parametrizes the 
effect of correlations. In this case the fluctuation variance becomes of the 
order of the proton mass at larger energies, namely $\sigma\simeq m_p$
at momentum of the order of

$$ p\simeq M_P \left (\frac{m_p}{M_P} \right )^{\frac{2}{3+\alpha}}~. $$

A detailed analysis of possible correlations
between fluctuations, namely an analytic determination of $\alpha$, is 
impossible at this stage because it implies a better knowledge of the theory,
and in particular of the dynamics of the QG regime.

Finally, a separate discussion is needed for those theories that 
include the relativity principle (exemplified by DSR models). The DSR 
theories are characterized by an extended Lorentz invariance \cite{dsr}
with two separate invariant scales: the light velocity and the Planck length.
Moreover, in the low energy limit of DSR, or for distances much larger 
than the Planck length, the usual Lorentz invariance is recovered. 

Using these two characteristics of the DSR theories it is easy to proove that 
particle kinematics in DSR is the same as in the usual Lorentz invariant 
theories. This result holds in the case in which there are no fluctuations
of energy and momentum. In the most general case in which fluctuations of 
energy and momentum are taken into account it is difficult to prove that the 
situation remains unchanged. Nevertheless, if in DSR the relativity principle 
remains at work also in the fluctuating case the DSR approach seems the most 
promising in order to escape the particles decays discussed in this paper 
that seems to invalidate all the other models.


\begin{thebibliography}{0}

\bibitem{kir}
D.A. Kirzhnits and V.A. Chechin, Sov. Jour. Nucl. Phys. {\bf 15}, 585 (1971).

\bibitem{lgm}
L. Gonzalez-Mestres, Proc. 26th ICRC (Salt Lake City, USA), {\bf 1}, 179 
(1999).

\bibitem{cam}
G. Amelino Camelia, J. Ellis, N.E. Mavromatos and S. Sarkar
Nature {\bf 393} (1998) 763.

\bibitem{colgla}
S. Coleman and S.L. Glashow, Phys. Rev. {\bf D59}, 116008 (1999).

\bibitem{noi}
R. Aloisio, P. Blasi, P.L. Ghia and A.F. Grillo,
Phys. Rev. {\bf D62}, 053010 (2000).

\bibitem{spain}
J.M. Carmona, J.L. Cortes, J. Gamboa and  F. Mendez, 
{\it preprint} hep-th/0301248. 

\bibitem{berto}
O. Bertolami and C.S. Carvalho, Phys. Rev. {\bf D61}, 103002 (2000);
O. Bertolami, {\it preprint} astro-ph/0012462.
 
\bibitem{gzk}
K. Greisen, Phys. Rev. Lett. {\bf 16}, 748 (1966);
G.T. Zatsepin and V.A. Kuzmin, Pis'ma Zh. Ekps. Teor. Fiz. {\bf 4}, 114  
(1966) [JETP Lett. {\bf 4}, 78 (1966)].

\bibitem{AGASA}
N. Hayashida {\it et al.} [AGASA collaboration], Phys. Rev. Lett. {\bf 73},
3491 (1994).

\bibitem{Hires} 
T. Abu-Zayyad {\it et al.} [Hires collaboration], astro-ph/0208243.

\bibitem{demarco}
D. De Marco, P. Blasi and A.V. Olinto,  
Astrop. Phys. {\bf 20}, 53 (2003).
 
\bibitem{Auger}
J. Bl\"umer {\it et al.} [Auger Collaboration], J. Phys. {\bf G29} 867 (2003).

\bibitem{EUSO}
EUSO space Observatory, http://www.euso-mission.org.

\bibitem{gamgam}
A.I. Nikishov, Sov. Phys. - JETP {\bf 14}, 393 (1962);
P. Goldreich and P. Morrison, Sov. Phys. - JETP {\bf 18}, 239 (1964);
R.J. Gould and G.P. Schreder, Phys. Rev. Lett. {\bf 16}, 252 (1966).

\bibitem{ford}
L.H. Ford Int. J. Theor. Phys. {\bf 38} 2941 (1999).

\bibitem{ng1}
Y.J. Ng, D.S. Lee, M.C. Oh and H. van Dam, 
Phys. Lett. {\bf B507} 236 (2001).

\bibitem{ng2}
Y.J. Ng, Int. J. Mod. Phys. {\bf D11} 1585 (2002) 
({\it preprint} gr-qc/0201022).

\bibitem{lieu}
R. Lieu, Astrophys. J. {\bf 568} L67 (2002).

\bibitem{noi2}
R. Aloisio, P. Blasi, A. Galante, P.L. Ghia and A.F. Grillo,
Astrop. Phys. {\bf 19}, 127 (2003).

\bibitem{gmestres}
L. Gonzalez-Mestres, {\it preprint} hep-ph/9905430.

\bibitem{liberati}
T. Jacobson, S. Liberati and D. Mattingly, 
Phys. Rev. {\bf D66}, 081302 (2002). 

\bibitem{agasa_a}
N.Hagashida {\it et al.}, Astrop. Phys. {\bf 10} 303 (1999).

\bibitem{fly_a}
D.J. Bird {\it et al.}, Ap. J. {\bf 511} 739 (1999).

\bibitem{agasa_f}
N. Sakay et al. Proceedings of 2001 ICRC.

\bibitem{Hires_a}
C.C.H. Jui {\it et al.}, Proceedings of 2001 ICRC.

\bibitem{blanton}
M. Blanton, P. Blasi and A.V. Olinto, Astrop. Phys. {\bf 15}, 275 (2001).

\bibitem{tom}
M. Jankiewicz, T.W. Kephart and T.J. Weiler, {\it preprint} hep-ph/0312221 

\bibitem{EGRET}
P. Sreekumar {\it et al.} [EGRET collaboration], 
Astroph. J. {\bf 494} (1998) 523.

\bibitem{GLAST}
GLAST Tlescope, http://www-glast.stanford.edu.

\bibitem{amepai}
G. Amelino-Camelia, Phys. Lett. {\bf B528} (2002) 181.

\bibitem{noi3}
R. Aloisio, P. Blasi, A. Galante and A.F. Grillo, 
Astrop. Phys. {\bf 20}, 369 (2003). 

\bibitem{dsr}
N.R. Bruno, G. Amelino-Camelia and J. Kowalski-Glikman, 
Phys. Lett. {\bf B522} 133 (2001). G. Amelino-Camelia, 
Int. J. Mod. Phys. {\bf D11} 1643 (2002). G. Amelino-Camelia
Nature {\bf 418} 34 (2002).

\bibitem{jap}
T. Tamaki, T. Harada, U. Miyamoto and T. Torii, 
Phys. Rev. {\bf D65} 083003 (2002).

\bibitem{camacho}
A. Camacho, Gen. Rel. Grav. {\bf 35} 319 (2003). 

\bibitem{kempf}
A. Kempf, G. Mangano and R.B. Mann, Phys. Rev. {\bf D52} 1108 (1995).


\end{thebibliography}
\end{document}